\documentclass[12pt,epsfig]{article}
\usepackage{amsmath}
\usepackage{graphicx}
\usepackage{slashed}

\def\s#1{\setbox0=\hbox{$#1$}%
\rlap{\ifdim\wd0>.7em\kern.22\wd0\else\kern.1\wd0\fi /}#1}

\newcommand{\beq}{\begin{equation}}
\newcommand{\eeq}{\end{equation}}
\newcommand{\bea}{\begin{eqnarray}}
\newcommand{\eea}{\end{eqnarray}}

\newcommand{\tto}{\!\to\!}

\newcommand{\gsim}{\lower.7ex\hbox{$
\;\stackrel{\textstyle>}{\sim}\;$}}
\newcommand{\lsim}{\lower.7ex\hbox{$
\;\stackrel{\textstyle<}{\sim}\;$}}

\renewcommand{\Im}{{\rm Im}\,}

\newcommand{\bibit}[1]{\bibitem{#1}}

\newcommand{\mhad}{\mu_{\rm hadr}}

\newcommand{\GeV}{\,\mbox{GeV}}
\newcommand{\MeV}{\,\mbox{MeV}}
\newcommand{\matel}[3]{\langle #1|#2|#3\rangle}

\newcommand{\eod}{\end{document}}

\newcommand{\msp}[1]{\mbox{\hspace*{#1mm}~}}

\begin{document}
\thispagestyle{empty}
\addtocounter{page}{-1}
\vspace*{-40pt}

\begin{flushright}
SI-HEP-2012-05

\end{flushright}
\vspace*{35pt}

\begin{center}
{\Large {\bf \hspace*{-1pt}Loop-Less Electric Dipole Moment 
of the Nucleon \\[2mm] in the Standard Model
}}\vspace*{50pt}

\vspace*{-30pt}

{\large \bf Thomas~Mannel ~and~ Nikolai~Uraltsev$\,^*$}\\
\vspace{15pt}
{\sl Theoretische Elementarteilchenphysik, Naturwiss.-Techn. Fakult\"at, \\
Universit\"at Siegen, 57068 Siegen, Germany}\\[15pt]

{\small {\sl$^*$ {\sf also} Department of Physics, University of Notre 
Dame du Lac,  Notre Dame, IN 46556, USA}\\[-1pt]
and \\[-1pt] 
{\sl St.\,Petersburg Nuclear Physics Institute, Gatchina,
St.\,Petersburg 188300, Russia}}\\

\normalsize

\vspace*{60pt}

{\bf Abstract}\vspace{-1.5pt}\\
\end{center}

\noindent
We point out that the electric dipole moment of the neutron in the Standard Model
is generated already at tree level to the second order in the weak
interactions due to bound-state effects, without short-distance Penguin 
loops. The related contribution has a regular nonvanishing chiral limit and
does not depend on the mass splitting between $s$ and $d$ quarks. We
estimate it to be roughly $10^{-31\,}\mbox{e$\cdot$cm}$ and expect a more accurate
evaluation in the future.
We comment on the connection between $d_n$ and the direct
CP-violation in $D$ decays. 

\newpage

Electric Dipole Moments (EDMs) of elementary particles and of composite
objects allow one to probe CP and T violation at the fundamental level. 
The Standard Model (SM) of particle physics has room for only two sources of 
CP violation. The first one roots in the strong
interaction proper, described by the so called $\vartheta$-term in the QCD 
Lagrangian \cite{thooft}. Described by a dimensionless 
coupling $\vartheta$,  this term
is flavor-diagonal and induces a large electric dipole moment of the neutron 
$d_n\!\approx \!\vartheta \!\times\! 3 \!\cdot\!10^{-16} e\!\cdot \!\mbox{cm}$, 
which by far exceeds the
experimental limit unless $\vartheta$ is extremely small, the notorious strong-CP
problem. It is then natural to
assume that $\vartheta$ is zero or nearly vanishes due to a symmetry or a dynamic
mechanism \cite{PQ}. 

The second source of CP violation lies in the electroweak sector of the SM and
is expressed through the irreducible phase in the CKM matrix $V_{jk}$ 
describing the flavor-nondiagonal weak transition amplitudes between quarks:
\beq
\frac{G_F}{\sqrt{2}}{\cal L}_w =\frac{G_F}{\sqrt{2}}J_\mu^\dagger J^{\mu} 
\mbox{ ~with~ } J^\mu=\sum_{j,k=1}^3 V_{jk}
\:\bar{u}_j \Gamma^\mu d_k\,, \qquad 
\Gamma^\mu\!=\!\gamma^\mu(1\!-\!\gamma_5).
\label{107}
\eeq
As a result,
it primarily manifests itself as CP violation in decays of kaons, $B$ and,
potentially, $D$-mesons. The CKM parameterization of flavor dynamics 
in the SM has been
successfully tested in detail over the last ten years in $K$ and $B$ mesons, 
in particular through its CP violating thread.

Being rooted in the intricacies of flavor-changing transitions, the CKM-CP
violation leads to extremely small effects in flavor-diagonal amplitudes 
such as EDMs; therefore EDMs in general and the neutron EDM $d_n$ in
particular are a very sensitive probe of the underlying source of CP
violation. Yet calculating the observable effect for hadrons in terms of the
fundamental parameters of the SM is often a difficult task requiring control
of the complicated hadronic dynamics in the grossly nonperturbative regime.

Following the success of the quark-gluon picture of the hadronic world and of
the qualitative understanding of the properties of hadrons based on the
constituent quark model, the early estimates aimed at calculating the EDMs of
quarks as the source of the nucleon EDM. It turns out that for quarks the SM
prediction is particularly suppressed: it emerges first at the three-loop
level where an additional loop with at least a gluon must be included
\cite{shabalin}. On top of this, the quark EDM has to be proportional to the
quark mass; this yields an additional suppression for the light quarks.  The
same applies to the color dipole moments of quarks considered as the simplest
induced CP-odd strong force generated through weak interactions at small
distances.

It has been noted a while ago \cite{gavela} that the severe chiral suppression
intrinsic to $u$ and $d$ quarks can be vitiated in composite hadronic systems
like nucleons.
The transition dipole moments changing $d$-quark into $s$-quark,
electromagnetic or color, are suppressed by the strange quark mass $m_s$,
and such flavor-changing transition without a quark charge change are mediated
by the loop-induced short-distance renormalization of the bare weak
interaction at some level via the so-called Penguin diagrams \cite{penguins}. 
A similar mechanism is believed to dominate the nucleon
EDMs in spite of the suppression associated with the heavy quark loops, 
although it is notoriously difficult to account for the long-range part of the
strong interactions it depends upon \cite{pospritz}. 

In this Letter we point out that the SM in fact generates the nucleon EDMs
already without any loop corrections, at the tree level of the second-order
weak interaction Lagrangian. This contribution has no chiral suppression
neither vanishes if strange and down quarks become nearly degenerate in
mass. Being loop-free, the induced EDM does not suffer from typically small
numerically perturbative factors characteristic to Penguin-induced effects
in spite of the accompanying logarithmic factors.

The price to pay is that the effect is inversely proportional to the square of
the charm quark mass. Since the latter is not excessively large in the
hadronic mass scale, the related suppression comes out relatively mild and
this effect may well dominate $d_n$ in the SM. It is described by multiquark
effective operators involving simultaneously $u$, $d$ and $s$ quark fields.

The numeric estimate of $d_n$ mediated in this way is presently rather
uncertain depending on the corresponding nucleon matrix elements. We do not
even speculate on its sign. Applying the natural counting rules we expect 
$|d_n|$ between $10^{-31}$ and  $10^{-30} e\!\cdot \!\mbox{cm}$. At the same
time we expect that definite estimates can be elaborated in the future
owing to the special form the effective CP-odd operators assume in the CKM 
case.

Since the contribution we focus on does not require loop effects, in the
following we neglect the well known short-distance renormalization of the weak
interaction Lagrangians altogether. This simplification makes the reasoning
transparent and all the expressions compact. Including the gluon-mediated
short-distance effects is straightforward, and we briefly comment on this.

Let us note that the similar ideas regarding the role of long-distance effects
in $d_n$ were discussed in the 1980s by I.\,Khriplovich with collaborators
\cite{khrip} as an extension of the original proposal by Gavela {\it et al.}
\cite{gavela}, and then in the 1990s by I.\,Khriplovich and A.\,Vainshtein.
Unfortunately, the latter analysis was not published and its main message
remains largely unknown. Reportedly, the authors focused on developing a
consistent low-energy effective theory for the corresponding CP-odd effects,
and therefore the loop renormalization occupied an important place in the
analysis. We are grateful to A.\,Vainshtein for the information and for useful
comments.

\section{Effective low-energy Lagrangian}

Through the neutron EDM we are looking into the CP-odd amplitudes that are
flavor-diagonal in every quark flavor. 
We assume that the renormalized effective $\vartheta$-term vanishes exactly
and the only source of CP nonconservation is the general CKM $3\!\times \!3$
unitary mixing matrix $V_{ij}$ describing the interaction of $W$ bosons with
quarks. Any first-order term in the conventional four-fermion weak interaction
Lagrangian with $\Delta F\!=\!0$ is automatically CP-invariant due to its
Hermiticity. The observable CP-odd effects appear from the second order 
in ${\cal L}_w$ and thus are proportional to $G_F^2$, being embodied in 
\beq
{\cal L}_2= \frac{G_F^2}{2} \!\int {\rm d}^4x \: \mbox{$\frac{1}{2}$}
iT\,\{{\cal L}_w(x)\,{\cal L}_w(0)\}.
\label{108}
\eeq
The generalized GIM-CKM mechanism ensures that the CP-odd piece of 
${\cal L}_2$  is finite in the local four-fermion approximation. 

Descending to a low normalization point we first integrate out top quark and
at the second stage, below $m_b$ the bottom quark as well. At the tree level
integrating them out essentially means that all the terms containing $t$ or
$b$ fields are crossed out of ${\cal L}_w$.

Within this setup we have effectively nearly a two-family weak Lagrangian
\beq
{\cal L}_w = J_\mu^\dagger J^\mu \mbox{~ with ~}
J_\mu = V_{cs}\, \bar{c} \Gamma_\mu s + V_{cd}\, \bar{c}
\Gamma_\mu d  + 
V_{us}\, \bar{u} \Gamma_\mu s + V_{ud}\, \bar{u}
\Gamma_\mu d \,, 
\label{110}
\eeq
yet not quite so, since the four $V_{kl}$ do not form a unitary matrix; in
particular, it is not CP-invariant. The phases in the four CKM couplings
cannot all be removed simultaneously by a redefinition of the four quark
fields, as quantified by 
\beq
\Delta \equiv  \Im U_4= \Im V_{cs}^*V_{cd} V_{ud}^*V_{us} \,.
\label{112}
\eeq

${\cal L}_w$ above contains altogether $16$ terms bilinear in $V$ and $V^*$.
The product ${\cal L}_2$ of two ${\cal L}_w$ in Eq.~(\ref{108}) already
consists of $256$ terms.\footnote{This general reasoning does not depend on
the tree level assumption: we can exclude $V_{tj}$ and $V_{ib}$ from the
pair products using the CKM unitarity.} Most of them change quark flavors --
only $64$ are flavor-diagonal. Of these most still contain the two pairs of
complex conjugate CKM factors and therefore are automatically CP-even.  Hence
a handful of terms are relevant to flavor-diagonal CP: only two different
operators are not CP-invariant driven by the same CKM product $U_4$, plus
their Hermitian conjugated partners. These non-local $8$-quark operators
include both $q$ and $\bar{q}$ fields for each of the four quark flavors. This
is readily understood: the CP-odd invariant $\Delta$ (as well as CP-violation
altogether) vanishes wherever any single CKM matrix element becomes zero.

\begin{figure}[hhh]
\vspace*{-5pt}
 \begin{center}
\includegraphics[width=14.cm]{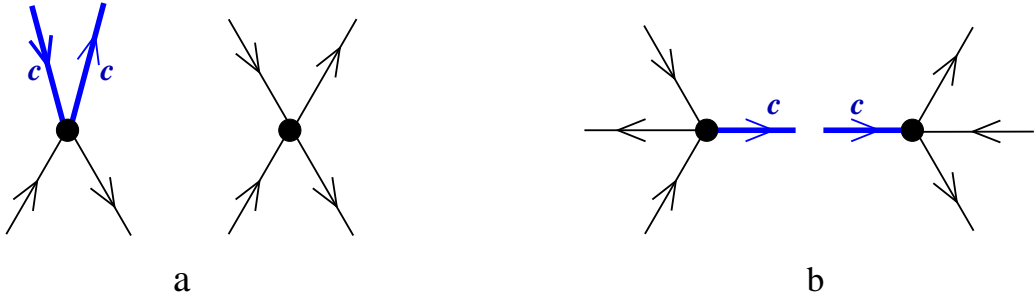}\vspace*{-25pt}
 \end{center}
\caption{ \small
Two types of CP-odd terms. Weak vertices must be off-diagonal in flavor, 
either for down-type (a) or up-type (b) quark. 
Solid dots denote the four-quark vertices. Light lines
correspond to $u$, $d$ or $s$ quarks, thicker lines stand for charm.
 }
\label{lwxlw}
\end{figure}

Being interested in the nucleon amplitudes, we need to eventually integrate
out the charm field as well.  Here the distinction between the above two terms
becomes important. If the two charm fields belong to the same four-fermion
vertex in the product Eq.~(\ref{108}) as in Fig.~\ref{lwxlw}a, they can be
contracted into the short-distance loop yielding, for instance, the usual
perturbative Penguins. These are the conventional source of the long-distance
CP-odd effects \cite{gavela,khrip}. The loop cannot be formed for the
alternative possibility where $c$ and $\bar c$ belong to different 
${\cal L}_w$, Fig.~\ref{lwxlw}b since the charm quark must propagate between the
two vertices. Such contributions therefore are routinely discarded.

\thispagestyle{plain}

On the contrary, our interest lies in the latter term: it does not involve
short-distance loops, and has a single charm propagator, 
although highly virtual in the hadronic scale. Each weak vertex contains a
flavorless  quark-antiquark pair, but these are light down-type quarks $d$ and
$s$ and are not contracted via a perturbative loop, instead going 
into the nucleon wavefunction. The corresponding operator is 
\beq
\frac{G_F^2}{2} \: V_{cs}V_{cd}^* V_{ud}V_{us}^* \int {\rm d}^4x\; 
iT\{(\bar{d}\Gamma_\mu c) (\bar{u}\Gamma_\mu d)_0 \cdot 
(\bar{c}\Gamma_\nu s)  (\bar{s}\Gamma_\nu u)_x \, +\, \mbox{H.c.}
\label{124}
\eeq
The Hermitian conjugate, apart from complex conjugation of the CKM product, is
simply the exchange between $s$ and $d$, $s\!\leftrightarrow\! d$. 

As the space separation $x$ in Eq.~(\ref{124}) is of order $1/m_c$,
eliminating charm results in a local OPE; the expansion parameter $\mhad/m_c$
is not too small and we need to keep a few first terms. The tree-level OPE is
particularly simple here and amounts to the series
\beq
c(0)\bar{c}(x)= \left(\frac{1}{m_c\!-\!i\!\not\!\!D}\right)_{0x}= 
\frac{1}{m_c} \delta^4(x) + \frac{1}{m_c^2}\delta^4(x)\, i\slashed{D} + 
\frac{1}{m_c^3}\delta^4(x)\, (i\slashed{D})^2 +...
\label{128}
\eeq
valid under the $T$-product. For purely left-handed weak currents in the SM
the odd powers of $1/m_c$ in Eq.~(\ref{128}) are projected out, including the
leading $1/m_c$ piece.  We then retain only the $1/m_c^2$ term and obtain the
local effective CP-odd Lagrangian
\bea
\label{130}
\tilde{\cal L}_- \msp{-4}&=&\msp{-4} -i 
\frac{G_F^2}{2m_c^2}\,\Delta \,(\tilde O_{uds}\!-\!\tilde O_{uds}^\dagger) , \\
\nonumber
\tilde O_{uds} \msp{-5}&=&\msp{-5} 
(\bar{u}\Gamma^\mu d)\, (\bar{d} \Gamma_\mu i\slashed{D}
\Gamma_\nu s)\,(\bar{s}\Gamma^\nu u) \!=\! 
(\bar{u}\Gamma^\mu d)\!\cdot\! 
\left[(\bar{d} \Gamma_\mu i\slashed{D}
\Gamma_\nu s)\!\cdot\!(\bar{s}\Gamma^\nu u) \!+\!  
(\bar{d} \Gamma_\mu i\gamma_\alpha
\Gamma_\nu s)\,i\partial^\alpha(\bar{s}\Gamma^\nu u)
\right]\!;
\eea
in the last expression the covariant derivative acts only on the
$s$-quark field immediately following it. 

To address the electric dipole moments we need to incorporate the electromagnetic
interaction. One photon source lies in the covariant derivative in the
operator $\tilde O_{uds}$, which includes electromagnetic potential along with
the gluon gauge field. It is proportional to the up-type quark electric charge
$+\frac{2}{3}$. The corresponding photon vertex is local and is given by the
Lorentz-vector six-quark operator which we denote as $O^\alpha_{uds}$:
\beq
O^\alpha_{uds}\!=\! (\bar{u}\gamma^\mu (1\!-\!\gamma_5)s)\, 
(\bar{s} \gamma_\mu i\gamma^\alpha
\gamma_\nu (1\!-\!\gamma_5) d)\,(\bar{d}\gamma^\nu (1\!-\!\gamma_5) u) .
\label{650}
\eeq
Another, non-local contribution is the $T$-product of the pure
QCD part of $\tilde O_{uds}$ 
\beq
O_{uds}\!=\! (\bar{u}\gamma^\mu (1\!-\!\gamma_5) s)\, 
(\bar{s} \gamma_\mu i\slashed{D}
\gamma_\nu (1\!-\!\gamma_5) d)\,(\bar{d}\gamma^\nu (1\!-\!\gamma_5) u) 
\label{651}
\eeq
with the light-quark electromagnetic 
current.\footnote{In general only the sum of the two terms yields the
transverse electromagnetic vertex; however, when projected on the dipole moment
Lorentz structures they separately conserve current.}
The total photon vertex is thus given by the effective CP-odd 
Lagrangian
\beq
 A_\alpha{\cal L}_-^\alpha \!=\! -e \,i\Delta \frac{G_F^2}{m_c^2} A_\alpha \!\left[
\mbox{$\frac{2}{3}$} O_{uds}^\alpha \!+ \!\int \!\!{\rm d}^4 x \:iT\{O_{uds}(0)\; 
J^\alpha_{\rm em}(x) \}
- \mbox{H.c.} \right]\!, 
\quad\: J^\mu_{\rm em}\!=\!\sum_q e_q \,\bar{q}\gamma^\mu q, 
\label{132}
\eeq 
where $A_\mu$ is the electromagnetic potential and $e$ is the unit charge. 

In principle, the local and non-local pieces above correspond to distinct
physics: one has photon emitted from distances of order $1/m_c$ while the
latter senses charge distribution over the $1/\mhad$ range.  The latter
usually dominates, however the specifics of the left-handed weak interactions
in the SM makes them of the same $1/m_c^2$ order.

An interesting feature of the considered contribution is that it remains
finite in the chiral limit and it does not vanish if $d$ and $s$ quarks become
nearly degenerate, at first glance contradicting 
the origin of the KM mechanism.\footnote{It would vanish if charm
and top become degenerate; considering the cases of degenerate bottom and
strange quarks, or charm and up makes no sense in this context since it has
been assumed as the starting point that $m_b, m_c \gg \mhad$ while $u$, $d$ and
$s$ are light quarks.} This in fact is fully consistent, since the external
state, the neutron, is explicitly $s\leftrightarrow d$ non-symmetric. This
highlights the difference with the short-distance effects for light quarks
where severe GIM-type suppression would arise and the EDM is
proportional to the powers of the light quark masses. 

The CP-odd operators contain strange quark fields. This means that the induced
effects would vanish in a valence approximation to nucleon where only $d$ and
$u$ quarks are active.  It is known, however, that even at low normalization
point the strange sea in nucleon is only moderately suppressed. The
large-$N_c$ perspective on the nucleons paralleling the picture of the baryon
as a quantized soliton of the pseudogoldstone meson field \cite{soliton} makes
this explicit: the weight of the operators with strange quarks in the chiral
limit is generally determined simply by the operator-specific Clebsh-Gordan
coefficients of the $SU(3)$ group.  This differs from the perturbative Penguin
effects which yield small coefficients whenever considered in the truly
short-distance regime.  The neutron dipole moment induced by the operator in
Eq.~(\ref{132}) does not need to vanish even in the quenched approximation to
QCD.

\section{Matrix elements}

The CP-odd operators $O^\alpha _{uds}\!-\!O^{\alpha\dagger} _{uds}$ and 
$O_{uds}\!-\!O^\dagger _{uds}$ have high dimension which is routinely regarded
as an evidence for being poorly defined for practical applications. However,
these particular operators possess very special symmetry properties, including
antisymmetry in respect to $s\leftrightarrow d$, which prohibit mixing with
lower-dimension operators, and make them a suitable object for the
full-fledged nonperturbative analysis.

The neutron EDM is obtained by evaluating the hadronic operator 
in Eq.~(\ref{132}) over the neutron
state. Since ${\cal L}_-^\alpha$ is T-odd, the matrix element vanishes for
zero momentum transfer and the linear in $q$ term 
describes $d_n$: 
\beq
\matel{n(p\!+\!q)}{{\cal L}_-^\mu}{n(p)}= -d_n\:
q_{\nu\,} \bar{u}_n(p\!+\!q)\sigma^{\mu\nu}\gamma_5 u_n(p).
\label{654}
\eeq
Neither of the matrix elements are easy to evaluate, although 
one may hope that just this particular contribution may eventually be determined
without major ambiguity, including the definitive prediction for the overall
sign. Clearly only the $P$-violating part of  $O^{(\alpha)}_{uds}$
contributes, but we keep them in the original form for the sake of explicit
symmetry and compactness. 

The contact operator $O^{\alpha}_{uds}$ is a product of three left-handed
flavor currents; $O_{uds}$ instead of the $\bar{s}d$ current has a flavor
non-diagonal left-handed partner of the quark energy-momentum tensor in the
chiral limit. Therefore it seems plausible that the required matrix elements
can be directly calculated within the frameworks like the Skyrme model
\cite{skyrme,soliton}, or in its dynamic QCD counterpart \cite{liquid} derived
in the large-$N_c$ limit from the instanton liquid approximation. This is the
subject of the ongoing study.

In the absence of better substantiated calculations we apply the simple
dimensional estimates to assess the expected size of $d_n$. 
We denote 
\beq
\matel{n(p\!+\!q)}{(\bar{u}_L\gamma^\alpha s_L)\, 
(\bar{s}_L \gamma_\alpha \gamma^\mu
\gamma_\beta d_L)\,(\bar{d}_L\gamma^\beta u_L) - (d\leftrightarrow s)} {n(p)} 
= -2i{\cal K}_{uds}\: q_\nu \bar{u}(p\!+\!q) \sigma^{\mu\nu}\!\gamma_5 u(p)
\label{660}
\eeq
for the local piece given by the operator $O_{uds}^\mu$.
The reduced matrix element ${\cal K}_{uds}$ has dimension of mass to the fifth 
power. We estimate it as
\beq
|{\cal K}_{uds}| \approx \kappa \;\mhad^5,
\label{662}
\eeq
where $\mhad$ is a typical hadronic momentum scale and $\kappa$ stands 
for the `strangeness suppression' to account for the fact that 
neutron has no valence strange quarks; $\kappa \!\approx \!1/3$ is taken as a
typical guess. 

The estimate for $d_n$ depends dramatically on the assumed value of
$\mhad$. It is known that the typical momentum of quarks in nucleon is around
$600\MeV$ or higher. Yet six powers of mass in Eq.~(\ref{662})
would come from the product of two local light quark currents each intrinsically
containing factors $N_c/8\pi^2$ when converted into the conventional momentum
representation. This is illustrated by the magnitude of the vacuum quark
condensate where such a factor effectively reduces $\mhad^3$ down to $\sim\!
(250\MeV)^3$. 

To account for this essential difference we assign a factor of
$(0.25\GeV)^3\!\equiv\!\mu_\psi^3 $ to each additional quark current in the
product, while the remaining dimension will be made of the powers of $\mhad$.
Then this contribution to $d_n$ becomes
\beq
|d_n| = \frac{32}{3} e \,\Delta \,\frac{G_F^2}{m_c^2}  |{\cal K}_{uds}|
\approx  3.3 \cdot 10^{-31} e\!\cdot \!\mbox{cm} 
\times \kappa \left(\frac{\mu_\psi}{0.25\GeV}\right)^6
\left(\frac{0.5\GeV}{\mhad}\right) \!,
\label{664}
\eeq
where $\Delta\!\simeq\!3.4\!\cdot\! 10^{-5}$ has been used. 
A potential enhancement may come from summation over the Lorentz indices 
in the currents.

The most naive estimates for the $d_n$ induced by the non-local
piece in Eq.~(\ref{132}) would yield a similar dimensional scaling 
except that no explicit
factor $e_c\!=\!\frac{2}{3}$ appears: the dimension of the non-local
$T$-product is the same as of $O_{uds}^\mu$ itself. Within the more careful
way advocated above the result is literally different:
\beq
|d_n|^{\rm n-loc} \approx e \,\Delta \,\frac{G_F^2}{m_c^2}  
32 \kappa\, \mu_\psi^9 \,\mhad^{-4}
\approx  1.2 \cdot 10^{-31} e\!\cdot \!\mbox{cm} 
\times \kappa \left(\frac{\mu_\psi}{0.25\GeV}\right)^9
\left(\frac{0.5\GeV}{\mhad}\right)^4\!,
\label{665}
\eeq
although numerically is not too far away. 

\begin{figure}[hhh]
\vspace*{-5pt}
 \begin{center}
\includegraphics[width=10.cm]{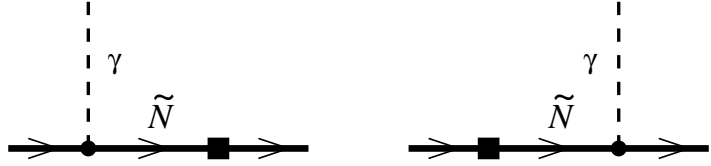}\vspace*{-18pt}
 \end{center}
\caption{ \small
Nonlocal contribution to $d_n$ with the intermediate $\tilde N$. Solid block
denotes the CP-odd part of the operator $O_{uds}$. 
}
\label{nonloc}
\end{figure}

Alternatively, to account for the specifics of the correlator with the
electromagnetic current in the non-local contribution we can consider the
contribution of the lowest resonant state, the $\frac{1}{2}^-$ nucleon
resonance $N(1535)$ referred to below as $\tilde N$. Denoting
\beq
\matel{n(p')}{J_{\rm em}^\mu (0)}{\tilde N (p)}\!=\! \rho_{\tilde N}
\bar{u}_n\gamma_5 \sigma^{\mu \nu} q_\nu  u_{\tilde{N}} , \quad 
\matel{\tilde N(p')}{O_{uds} (0)\!-\!O_{uds} (0)^\dagger}{n(p)}\!=\! 16 i
{\cal N}_{\!uds}\bar{u}_{\tilde{N}} u_n
\label{666}
\eeq
we obtain for the Feynman diagrams in Fig.~\ref{nonloc} 
\beq
d_n^{(\tilde N)} = -e \: \Delta \frac{32 G_F^2}{m_c^2} 
\left(\frac{  \rho_{\tilde N}\, {\cal N}_{\!uds}}
{M_{\tilde{N}}\!-\! M_N} \right) .
\label{668}
\eeq

\thispagestyle{plain}

The operator $O_{uds}$ has dimension ten, and the similar
dimensional estimate reads
\beq
|{\cal N}_{\!uds}| \approx \kappa \;\mu_\psi^6 \,\mhad.
\label{670}
\eeq
$\rho_{\tilde N}$ can be estimated from the measured transition 
$ \tilde{N} \tto n\!+\!\gamma$  approximating the rate with the dipole expression
\beq
\Gamma (\tilde{N} \to n \gamma) \simeq \frac{1}{2}\alpha_{\rm em}  
\rho_{\tilde N}^2 M_{\tilde{N}}^3 
\left(\!1\!-\!\frac{M_n^2}{M_{\tilde{N}}^2} \!\right)^{\!3}
\approx (380 \pm 180) \, \mbox{keV} 
\label{672}
\eeq
yielding $\rho_{\tilde N} \!\approx\! (0.34 \!\pm\! 0.08) \, \GeV^{-1}$.
Finally this estimate would read 
\beq
|d_{n}|^{(\tilde N)} \!\approx \! e \, \Delta \frac{32 G_F^2}{m_c^2}  
\kappa\,\mu_\psi^6 \,\mhad \, \frac{\rho_{\tilde N} }{M_{\tilde{N}}\!-\! M_n} 
\approx \!
1.4 \cdot 
10^{-31} \mbox{e$\cdot$cm} \times \kappa \left( \frac{\mu_\psi}{\mbox{0.25\GeV}}
\right)^6 \!\left( \!\frac{\mhad}{\mbox{0.5\GeV}} \!\right)
. 
\label{674}
\eeq
This value appears quite consistent with the direct dimensional estimate of
the non-local contribution, in particular considering 
the fact that the lowest excited state alone may not necessary saturate it. 

Therefore, our estimate for $d_n$ in the SM literally centers around 
$10^{-31\,}\mbox{e$\cdot$cm}$ although even the values $5$ to $10$ times larger
may not be excluded. 

\section{Discussions and conclusions}

So far we have neglected short-distance loop effects. They can be
included in the standard way. Importantly, the gluon
corrections do not change the symmetry properties and therefore 
cannot radically modify the structure of the result. 

The evident impact is renormalization of the individual ${\cal L}_w$
modifying the strength and inducing different color flow. This generates
different flavor allocation between the brackets in the operators $O_{uds}$
and $O^\alpha_{uds}$, yet does not change their properties. In particular, the
overall antisymmetry with respect to $s$ and $d$ is never
modified. Qualitative difference neither arises from the Penguin-induced 
operators with right-handed currents. Most of these changes affect the
conventional contributions belonging to Fig.~\ref{lwxlw}a. 

It turns out the situation neither changes where an external magnetic photon
or gluon attaches to the $\bar{Q}Q$ loop from a single ${\cal L}_w$: the sum
of all the diagrams vanishes or is proportional to the external light quark
field mass. In any case such renormalization emerges only at the two-loop
level and apparently is too suppressed numerically.

As a result, the short-distance loops should not radically affect $d_n$
obtained already at the tree level, unless the latter happens to suffer from
accidental numeric cancellations. All the potential corrections have
counterparts in the gluon-induced renormalization of the operators $O_{uds}$
and $O^\alpha_{uds}$ themselves.  Therefore we may a priori expect a moderate
overall enhancement of $d_n$ due to increased size of $c_-$ down from the
$W$-boson scale; the final conclusion would require evaluating the matrix
elements accounting, in particular, for the alternative color contractions in
the operators.  \vspace*{3pt}

To summarize, we have described the contribution to the EDM of nucleons
proportional to $\Delta G_F^2 \mhad^5/m_c^2$ that does not require loop
effects associated with heavy flavors or short-distance gluon corrections. It
is finite in the chiral limit for light quarks and would not vanish even if
$m_s$ and $m_d$ were very close.  Free from typically small perturbative loop
factors it plausibly dominates $d_n$ in the SM, yet it is interesting
regardless of the precise magnitude.

Within the effective theory describing the SM below the bottom quark scale,
charm quark must necessarily appear for the CP-violation to show up.
Therefore, if we avoid perturbative charm Penguin loops, they are replaced by
the terms suppressed by powers of $1/m_c$. We had encountered similar
nonperturbative charm effects dubbed `Intrinsic charm' in the inclusive beauty
decays \cite{IC}; they were distinct, yet not too significant and mattered
only as long as high precision was aimed at.

The situation is different for $d_n$ and evidently in an important aspect:
here we deal with `tree' nonperturbative charm and it is suppressed by powers
of $1/m_c$ vs.\ powers of $1/(2m_c)$ for charm Penguins.  Moreover, the
absence of the loop factor here apparently leads to a notably milder
suppression. For instance, no $\alpha_s(2m_c)$ enters as a prefactor.

The left-handed structure of the charged currents in the SM yields the second
power of the charm mass suppression and, when considering the photon
interaction, the local and the non-local contribution of the same order. In
this situation the local piece possibly dominates; this can be verified having
more elaborate estimates of the nucleon matrix elements. Although rather
uncertain at the moment, we believe the proposed contribution to the $d_n$ in
the SM can eventually be evaluated without a too significant uncertainty,
owing to a rather peculiar form of the effective CP-odd operator.

The chiral properties of the SM forbid the effective CP-violating scalar
pion-to-nucleon coupling at small momenta, which is often important for the
atomic EDMs. This applies as well to the terms power-suppressed in $1/m_c$.

Since any CKM-generated contribution to $d_n$ must be proportional to $\Delta
G_F^2$, the considered effect is not too far from the top benchmark $\Delta
G_F^2 \mhad^3$: it is moderated by $1/m_c^2$, and may be somewhat numerically
suppressed through explicitly containing the strange quark field, a feature
likewise unavoidable for CKM. It also does not generate chiral enhancement
$\ln{\mhad^2/m_\pi^2}$ which would be possible with the right-handed currents.
Therefore we conclude that $d_n$ in the SM is moderately suppressed compared
to the intrinsic smallness of the flavor-diagonal CP violation built in the
CKM ansatz.

The discussed mechanism generating $d_n$ roots in the CP-odd interference 
of the $\Delta C\!=\!1$, $\Delta S\!=\!0$ amplitudes in the usual weak 
decay Lagrangian. This is precisely what induces the mixing-free `direct' CP
asymmetry in $D$ decays. Therefore, New Physics (NP) phenomena in the latter would
directly affect $d_n$ at the stated level \cite{edm2}. It turns out that,
barring contrived cancellations, the NP $d_n$ enhancement is greater than its
effect on the $D$-decay asymmetries. This indicates that, in a certain sense,
the particular SM implementation of the CKM ansatz nearly minimizes 
the intrinsic size of $d_n$.

The increase in the EDM depends on the details of the underlying NP
interaction. For instance, the amplitudes with the right-handed charm field
lead to $1/m_c$ scaling of $d_n$, and the right-handed light-quark currents
tend to generate an overall enhancement. The latter also induce the CP-odd
$\pi NN$ coupling. Taking the reported asymmetry in $D\tto K^+K^-$ and $D\tto
\pi^+\pi^-$ at face value \cite{LHCb} we typically find an enhancement in
$d_n$ by a factor of $10$ to $100$, with the case where the NP amplitude is
described by $\bar{c} \sigma_{\mu\nu} F^{\mu\nu}\gamma_5 u$ alone leading to
$d_n$ up to $5 \!\cdot\! 10^{-27}\mbox{e$\cdot$cm}$. Such an extreme
possibility does not look too natural from the perspective of speculating
about the underlying flavor theory, though.  
\vspace*{5pt}

\noindent
{\bf Acknowledgments:} 
~We are grateful to I.~Bigi and A.~Khodjamirian for comments. N.U.\ is happy
to thank V.~Petrov for invaluable discussions of the chiral baryon properties.
Our special thanks go to A.~Vainshtein for numerous discussions and insights.
The work was largely supported by the German research foundation DFG under
contract MA1187/10-1 and by the German Ministry of Research (BMBF), contracts
05H09PSF; it also enjoyed a partial support from the NSF grant PHY-0807959 and
from the grant RSGSS 4801.2012.2.

\end{document}